\documentclass[12pt]{iopart}
\usepackage[colorlinks=true,citecolor=blue,linkcolor=blue,urlcolor=blue]{hyperref}
\usepackage{amssymb}
\usepackage{graphicx}
\usepackage[compress]{cite}
\usepackage[usenames]{color}
\usepackage{subfigure}
\usepackage{epsfig}

\newcommand{\ie}{{\it{i.e.~}}}
\newcommand{\red}[1]{\textcolor{red}{#1}}
\newcommand{\ket}[1]{\vert #1 \rangle}

\begin{document}

\title{Quantum Pattern Recognition in Photonic Circuits}

\author{Rui Wang$^{1}$, Carlos Hernani-Morales$^{2}$, Jos\'e D. Mart\'in-Guerrero$^{2}$, Enrique Solano$^{1,3,4,5}$, Francisco Albarr\'an-Arriagada$^{1}$}

\address{$^{1}$ International Center of Quantum Artificial Intelligence for Science and Technology (QuArtist) and Department of Physics, Shanghai University, 200444 Shanghai, China}
\address{$^{2}$ IDAL, Electronic Engineering Department, University of Valencia, Avgda. Universitat s/n, 46100 Burjassot, Valencia, Spain}
\address{$^{3}$ Department of Physical Chemistry, University of the Basque Country UPV/EHU, Apartado 644, 48080 Bilbao, Spain}
\address{$^{4}$ IKERBASQUE, Basque Foundation for Science, Plaza Euskadi 5, 48009 Bilbao, Spain}
\address{$^{5}$ Kipu Quantum, Kurwenalstrasse 1, 80804 Munich, Germany}

\ead{\mailto{enr.solano@gmail.com}, \mailto{pancho.albarran@gmail.com},\mailto{jose.d.martin@uv.es}}

\begin{abstract}

This paper proposes a machine learning method to characterize photonic states via a simple optical circuit and data processing of photon number distributions,  such as photonic patterns. The input states consist of two coherent states used as references and a two-mode unknown state to be studied. We successfully trained supervised learning algorithms that can predict the degree of entanglement in the two-mode state as well as perform the full tomography of one photonic mode,  obtaining satisfactory values in the considered regression metrics.
		
\end{abstract}

\section{Introduction}

Quantum information processing deals with the manipulation of quantum states in order to perform quantum informational tasks such as quantum algorithms~\cite{Montanaro2016NPJ}, quantum error correction~\cite{Roffe2019ContempPhys}, quantum cryptography~\cite{Scarani2009RMP},  or quantum teleportation~\cite{Pirandola2015NatPhotonics}. It is known that quantum information has the potential to outperform classical information protocols~\cite{Bravyi2020NatPhys,Arute2019Nature,Zhong2020Science,Yin2020Nature,Wu2021arXiv}. However,  since a quantum system is modified after measurement, extracting arbitrary information from a quantum state may require many copies. In general,  quantum tomography (QT) can be employed for the full reconstruction of a quantum state or a quantum operator. In a nutshell, QT implies the measurement of the expectation values of several operators or the use of a mutually unbiased basis~\cite{Hradil1997PRA,DAriano2001PRL,Adamson2010PRL}, which is in general a hard experimental task. In the last years,  some efficient protocols have been proposed assisted by machine learning (ML) algorithms, although it may still be tricky depending on the dimension of the quantum system~\cite{Tiunov2020Optica,Palmieri2020npjQI, Torlai2018NatPhys,Lohani2020MLST,AlbarranArriagada2018PRA,Yu2019AQT}. 
	
Quantum information can be encoded, decoded, and manipulated in a variety of physical systems, for example, photonic systems~\cite{Flamini2018RPP}, solid state devices \cite{Yao2012NatComm}, trapped ions~\cite{Bruzewicz2019APR,Eltony2016QIP}, and superconducting architectures~\cite{Devoret2013Science,Wendin2017RPP}. Photonic platforms present several advantages for quantum information protocols,  such as long coherence times and full connectivity, allowing long-distance quantum communication and quantum key distribution, among many other achievements~\cite{Liao2017Nature,Ren2017Nature,Chen2021Nature}. Different photonic degrees of freedom, including polarization, spectral, spatial, and temporal modes can be used to encode information~\cite{Flamini2018RPP}, providing a huge variety of experimental resources for many quantum information tasks.
	
One of the most intriguing experimental resources for photonic platform is boson sampling, which is a model of non-universal quantum computation. It consists in  measuring or sampling the quantity of photons in a distribution produced by a linear interference device given an initial photonic Fock state~\cite{Aaronson2011ACM}. It is a complex problem that cannot be efficiently simulated on conventional computers. However,  with accessible experimental requirements,  it can be simulated on linear optical quantum computers. It is attractive that the experimental setup of boson sampling only requires single-photon sources, photodetectors, and linear optical elements, \ie beam splitters and phase shifters~\cite{Gard2015arxiv}. Such a feasibility has encouraged and inspired many research teams for lab implementations~\cite{Broome2013Science, Tillmann2013NP, Bentivegna2015SA,Spagnolo2014NP,Wang2019PRL}. Particularly, recent experiments have shown the potential of photonic technologies to obtain quantum advantage for scientific relevant problems; for instance,  quantum supremacy of Gaussian boson sampling with 76 photons~\cite{Zhong2020Science}, and the so-called timestamp boson sampling, that brings to light the potential of circuits to get memory effects, thus paving the way for the implementation of neuromorphic quantum computing~\cite{Zhou2020arXiv,Gao2020arXiv}.

On a separate note, it is known that ML algorithms help to extract information from large and complex data sets, encompassing different techniques with sound mathematical grounds~\cite{Alpaydin2004MIT, Shwartz2014CUP}. The information is extracted by means of learning algorithms,  that teach models which usually have no a priori knowledge of the problem.  The increasing availability and size of data sets has spread the use of ML~\cite{Mathur2019Apress, Olivas2009IG}, oftentimes involving reliable applications at academic and commercial levels. 

The main goal of this manuscript is to show a simple and experimentally feasible photonic architecture capable to produce photonic patterns which assisted by ML algorithms could give an estimation about quantum features hard to calculate experimentally. To this end, we consider whether the full tomography of a photonic mode and two-mode entanglement can be extracted via sampling the output distribution of photons. In order to obtain a relation between the distribution probability of the output state and the information of the unknown state, we propose a particular linear optical circuit with four spatial modes, and calculate the corresponding permanent of the submatrix of the unitary matrix. We make use of a data set of probability patterns, that corresponds with the photon number distributions of output states. Then, we change the parameters of the unknown state to build a supervised learning algorithm for estimating the state of a new probability pattern, distinct from the training one.
	
The rest of the article is organized as follows. In section~\ref{Sec2}, we briefly review the most important aspects of boson sampling. Sections~\ref{Sec3} and~\ref{Sec4} introduce our four-mode optical circuit for state estimation of a photonic-mode, and entanglement estimation for a bipartite system, respectively. The results of the ML approach for state estimation and entanglement estimation are shown in section~\ref{Sec5}, ending up the paper with the conclusions of the work in section~\ref{Sec6}.

\section{\label{Sec2}Boson-sampling model}

A boson-sampling experiment consists in measuring the photon number probability produced by the interference of $N$ photonic states, usually indistinguishable single-photon states, via an $M$-mode linear network. The distribution can be obtained by computing permanents of the submatrix derived from the unitary transformation matrix of the network~\cite{Scheel2004arxiv}. The calculation of the permanent for an $n\times n$ complex matrix is a $\#P$-hard task for classical computers and $\#P$-complete for a $(0,1)$-matrix~\cite{Valiant1979TCS}, which means that the simulation of a boson-sampling experiment by classical devices is inefficient.

Without loss of generality, the output state of an $M$-mode optical circuit after the interference of an $N$-photon state can be calculated as follows. First, the input state is given by
\begin{eqnarray}
|\psi_{in}\rangle && = \ket{I_1,I_2,...,,I_M} = \left(\prod_k\frac{\hat{a}_k^{\dagger I_k}}{\sqrt{I_k!}}\right)\ket{0} \, ,
\end{eqnarray}
where $\hat{a}_k^{\dagger}$ is the photon creation operator for the $k^{th}$ mode, $I_k$ is the number of photons in the $k^{th}$ mode, and $\sum_k I_k = N$. The input and output states are related by a unitary transformation,
\begin{equation}
\ket{\psi_{out}}=\mathbb{U}\ket{\psi_{in}}=\mathbb{U}\left(\prod_k\frac{\hat{a}_k^{\dagger I_k}}{\sqrt{I_k!}}\right)\mathbb{U}^{\dagger}\mathbb{U}\ket{0} \, ,
\end{equation}
where $\mathbb{U}\ket{0}=\ket{0}$ because a linear optical circuit preserves the photon number. Now, the operator transformation reads
\begin{equation}
\mathbb{U}\hat{a}_k^{\dagger}\mathbb{U}^{\dagger} = \sum_j u_{k,j} \, a_j^{\dagger} \, .
\end{equation}
Again,  as a linear optical circuit preserves the number of photons, then $u_{k,j}$ defines a unitary matrix $U$, which represents a superoperator that acts over the space of the creation operators. Then, the output state reads
\begin{eqnarray}
\ket{\psi_{out}}&&= \left[\prod_k\frac{(U\hat{a}_k^{\dagger })^{I_k}}{\sqrt{I_k!}}\right]\ket{0}=\sum_O \gamma_O\bigotimes_k\ket{O_k}=\sum_O \gamma_O\ket{\psi_O} \, ,
\end{eqnarray}
where $O$ is a photon-mode configuration containing $N$ photons, $\gamma_O$ is the superposition factor of the configuration $O$, and $\ket{O_k}$ is the photon number state for the $k^{th}$-output mode in the $O^{th}$ configuration. The probability of measuring configuration $O$ is given by $P_O = |\gamma_O|^2$. The probability $P_O$ is given by
\begin{equation}
P_O=\frac{\left|Per[\Lambda_{|\psi_I\rangle,|\psi_O\rangle}]\right|^2}{I_1!...I_M!O_1!...O_M!} \, ,
\label{prob-def} 
\end{equation}
where $Per[\cdot]$ is the permanent and $\Lambda_{|\psi_I\rangle,|\psi_O\rangle}$ is an $N\times N$ matrix, which can be obtained from the elements $u_{jk}$ of  that defines $U$~\cite{Gard2015arxiv}. The matrix $\Lambda_{|\psi_I\rangle,|\psi_O\rangle}$ reads
\begin{eqnarray}
(\Lambda_{|\psi_I\rangle,|\psi_O\rangle})_{j,k}=U_{p,q}=u_{p,q} &\iff &S^O_{p-1}+1\le j\le S^O_p \, , \nonumber\\
&& S^I_{q-1}+1\le k\le S^I_q \, ,
\end{eqnarray}
where
\begin{eqnarray}
S^O_{\ell}=\sum_{i=1}^\ell O_i \quad \land \quad S^I_{\ell}=\sum_{i=1}^\ell I_i 
\end{eqnarray}
are the number of output and input photons until mode $\ell$, respectively,  being $S_0^{I(O)}=0$. It should be borne in mind that, for any linear network with $M$ modes, we have that $S^I_M=S^O_M = N$. For example, if $\ket{\psi_I}=\ket{0_11_20_32_4}$ and $\ket{\psi_O}=\ket{1_10_21_31_4}$, then $S^I_1=0$, $S^I_2=1$, $S^I_3=1$, and $S^I_4=3$, while $S^O_1=1$, $S^O_2=1$, $S^O_3=2$, and $S^O_4=3$. Then, the matrix $\Lambda_{|\psi_I\rangle,|\psi_O\rangle}$ reads
\begin{eqnarray}
\Lambda_{|\psi_I\rangle,|\psi_O\rangle}=\left(
\begin{array}{ccc}
 U_{1,2} & U_{3,2} & U_{4,2} \\
U_{1,4} & U_{3,4} & U_{4,4} \\
U_{1,4} & U_{3,4} & U_{4,4}     
\end{array}
\right)
\end{eqnarray}
	
\begin{figure}[t]
\centering
\includegraphics[height=4cm,width=8cm]{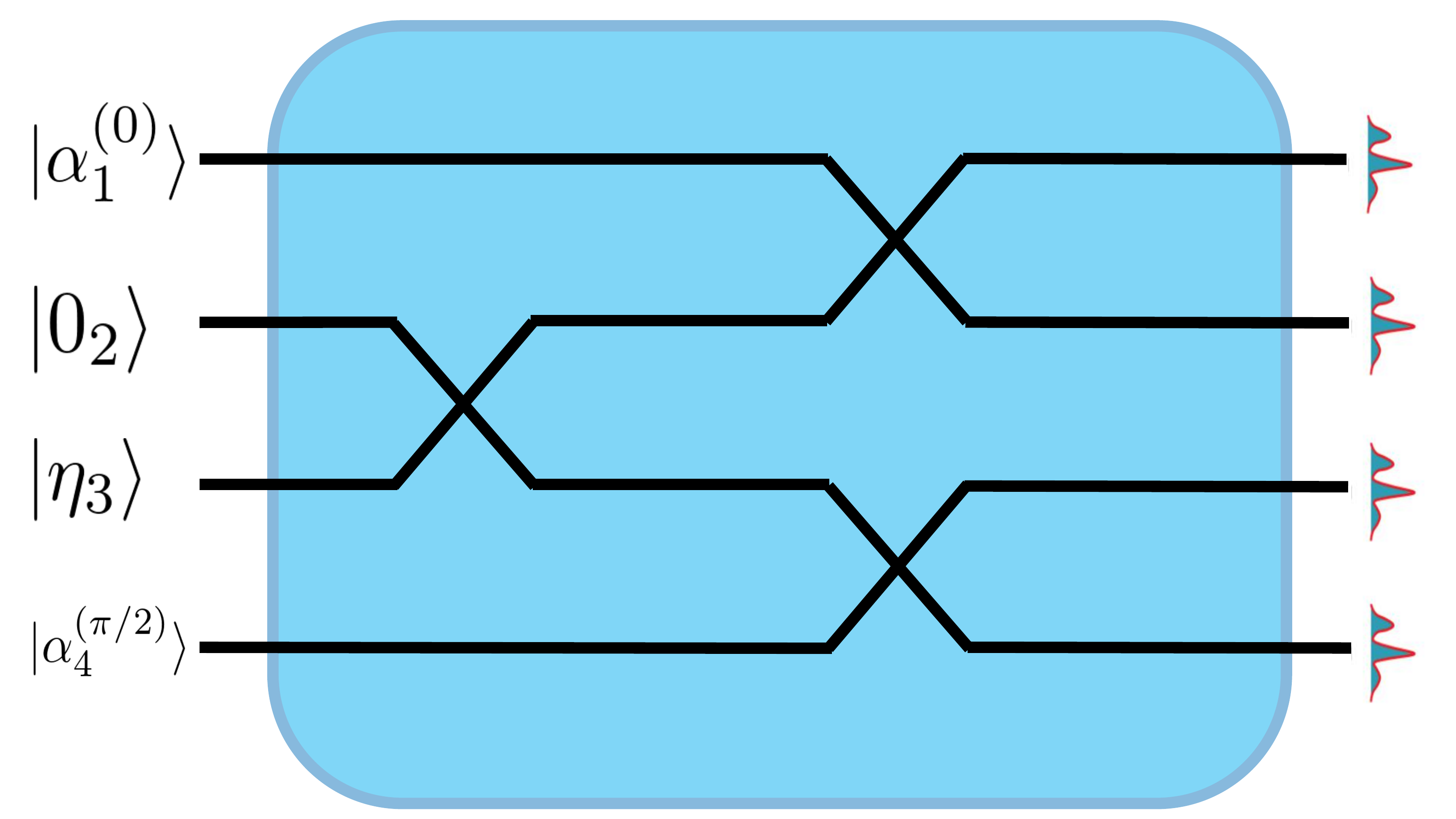}
\caption{Core optical circuit for quantum tomography made up of three beam splitters and the initial state $|\psi_{I}\rangle=|{\psi^{(0)}_1}\rangle |0_2\rangle |\eta_3\rangle |{\psi^{(\pi/2)}_4}\rangle$.}
\label{Fig01}
\end{figure}

\section{\label{Sec3}Optical circuit for state tomography}
We used a boson-sampling circuit to achieve the full quantum tomography of an unknown superposition state in the Fock basis. For obtaining more information of the unknown state, we proposed a simple optical circuit formed by three $50\%-50\%$ beam splitters, as shown in Fig.~\ref{Fig01}. The unitary matrix $U$ for this circuit is given by
\begin{eqnarray}
\hat{U}=\left(
\begin{array}{cccc}
a&-a^2&a^2&0\\
a&a^2&-a^2&0\\
0&a^2&a^2&-a\\
0&a^2&a^2&a\\
\end{array}
\right)\end{eqnarray}
where $a=1/\sqrt{2}$. The input states for the first and the last mode are coherent states given by
\begin{equation}
|{\alpha_{j}^{(\theta)}}\rangle = \sum_{n}^{\infty}\frac{e^{-\frac{1}{2}|\alpha|^2} e^{in\theta}|\alpha|^n}{\sqrt{n!}}|n_j\rangle \, ,
\end{equation}
 with $\theta = 0$ for the mode $1$ and $\theta=\pi/2$ for the mode $4$. The initial state for mode $2$ is the vacuum state $\ket{0_2}$, and for the mode $3$ it is an arbitrary unknown state $|\eta_3\rangle = \sum_{\ell = 0}^{N}r_\ell e^{i\phi_\ell}|n_3\rangle$.
 
 The considered initial state of our optical circuit is given by 
\begin{equation}
\ket{\Psi}=\sum_{m,n=0}^{\infty} \sum_{\ell=0}^{N}i^n\frac{e^{-|\alpha|^2} |\alpha|^{m+n}}{\sqrt{m!n!}}r_\ell e^{i\phi_\ell}\ket{m_1}\ket{0_2}\ket{\ell_3}\ket{n_4} \, .
\end{equation}
This state is a superposition of different four-mode states with different number of photons. For the sake of clarity and simplicity, we rewrite the initial state as
\begin{eqnarray}
\ket{\Psi}=\sum_{s=0}^{\infty}|\psi^I_{s}\rangle \, ,
\end{eqnarray}
where $|\psi^I_{s}\rangle$ is a non-normalized state with the superposition of all the four-mode states with $s$ photons, which reads
\begin{eqnarray}
|\psi_{s}^I\rangle = \sum_{\ell=0}^N \sum_{n=0}^{s-\ell} i^n\frac{e^{-|\alpha|^2} |\alpha|^{s-\ell}}{\sqrt{(s-\ell-n)!n!}}r_\ell e^{i\phi_\ell}\times\ket{(s-\ell-n)_1}\ket{0_2}\ket{\ell _3}\ket{n_4} \, .
\end{eqnarray}
Then, the output state is a superposition of different configurations of how the photons may have arrived to their modes,
\begin{eqnarray}
|\psi_{s}^O\rangle=\sum_{C_s}\gamma_{C_s}|ghkf\rangle_{C_s},
\end{eqnarray}
where $C_s$ denotes a particular configuration with $s$ photons. The probability of measuring a configuration $C_s$ is given by
\begin{eqnarray}
P_{C_s}=e^{-2|\alpha|^2}\left|\sum_{\ell=0}^{N_{min}} \sum_{n=0}^{s-\ell}i^n \frac{r_\ell e^{i\phi_\ell}|\alpha|^{s-\ell} Per [\Lambda_{|\psi_I\rangle,|\psi_O\rangle}] }{(s-\ell-n)!n!\sqrt{\ell!g!h!k!f!}}\right|^2,
\label{def_prob}
\end{eqnarray}
where $N_{min}=\textrm{min}(s,N)$. Next, we will consider particular cases of the maximum number of photons of the state $\ket{\eta_3}$ ($N=1$ and $N=2$), and then we will extend them to an arbitrary superposition of Fock states.

\subsection{\label{sec:level3} The case N=1}

The unknown state is $|\eta_3\rangle=r_0|0\rangle+r_1 e^{i\phi_1}|1\rangle$ when $N=1$. Firstly, for $s=0$ we have only one possible output $C_0=\ket{0000}$ with probability given by Eq.~(\ref{def_prob}),
\begin{equation}
P_{C_0}=e^{-2|\alpha|^2}r^2_0
\end{equation}
where we consider $\phi_0=0$. For $s=1$ we have four different outputs, $C_1\in\{\ket{1000},\ket{0100},\ket{0010},\ket{0001}\}$, with the general formula to get the probability given by
\begin{eqnarray}
P_{C_1} = e^{-2|\alpha|^2}\Big|&&|\alpha|r_0(Per[\Lambda_{(\ket{1000},\ket{C_1})}]+iPer[\Lambda[_{(\ket{0001}\ket{C_1},])})\nonumber\\
&&+r_1e^{i\phi_1}Per[\Lambda[_{(\ket{0010},\ket{C_1})}]\Big|^2.
\label{prob_n1}
\end{eqnarray}

The probabilities are
\begin{eqnarray}
&&P_{\ket{1000}} = \frac{1}{2}e^{-2|\alpha|^2}\left(|\alpha|^2r_0^2+\frac{1}{2}r_1^2+\sqrt{2}|\alpha|r_0r_1\cos\phi_1\right) \, , \\
&&P_{\ket{0100}} =  \frac{1}{2}e^{-2|\alpha|^2}\left(|\alpha|^2r_0^2+\frac{1}{2}r_1^2-\sqrt{2}|\alpha|r_0r_1\cos\phi_1\right) \, , \\
&&P_{\ket{0010}} =\frac{1}{2}e^{-2|\alpha|^2}\left(|\alpha|^2r_0^2+\frac{1}{2}r_1^2-\sqrt{2}|\alpha|r_0r_1\sin\phi_1\right) \, , \\
&&P_{\ket{0001}} =\frac{1}{2}e^{-2|\alpha|^2}\left(|\alpha|^2r_0^2+\frac{1}{2}r_1^2+\sqrt{2}|\alpha|r_0r_1\sin\phi_1\right) \, .
\end{eqnarray}
Note that the parameters $r_0$, $r_1$, and $\phi_1$ are encoded in the probabilities.  Therefore,  these parameters can be obtained by calculating the sum or difference of the probability distributions.

\subsection{\label{sec:level3} The case N=2}

In this case, we consider $|\eta_3\rangle=r_0|0\rangle+r_1 e^{i\phi_1}|1\rangle+r_2e^{i\phi_2}|2\rangle$\red{;} the probabilities for $s<2$ are the same as in the previous case. For $s=2$, we have
\begin{eqnarray}
P_{C_2}&&=\frac{e^{-2|\alpha|^2}}{g!h!k!f!}\Big[\frac{|\alpha|^2r_0}{2}\big(Per[\Lambda_{(\ket{2000},\ket{C_2})}]\nonumber\\
&&-Per[\Lambda_{(\ket{0002},\ket{C_2})}]+2iPer[\Lambda_{(\ket{1001},\ket{C_2})}]\big)\nonumber\\
&&+|\alpha|r_1e^{\i\phi_1}\big(Per[\Lambda_{(\ket{1010},\ket{C_2})}]+iPer[\Lambda_{(\ket{0011},\ket{C_2})}]\nonumber\\
&&+\frac{r_2e^{\phi_2}}{\sqrt{2}}Per[\Lambda_{(\ket{0020},\ket{C_2})}]\big)\Big]^2 \, .
\end{eqnarray}

For $\ell = 0$ and $g+h=2-n$, the permanent is given by 
\begin{equation}
Per[\Lambda_{\ket{(2-n)_1}\ket{0_2}\ket{0 _3}\ket{n_4},|C_2\rangle]
}= (-1)^ka^2(2-n)!n!
\end{equation}

For $\ell \neq 0 $, we have the case $g+h-d=2-\ell-n$ and $k+f-q=n$, where $d+q=\ell$. Here, the corresponding permanent reads
\begin{eqnarray}
&& Per[\Lambda_{\ket{(2-\ell-n)_1}\ket{0_2}\ket{\ell _3}\ket{n_4},|C_2\rangle}]\nonumber\\
 = && B_\ell^d \cdot a^{2+\ell}\cdot \Big[\sum_{x=0}^d B_d^x(-1)^x(2-\ell-n)!\frac{g!}{[g-(d-x)]!}\frac{h!}{(h-x)!}\Big] \cdot \nonumber \\
&& \Big[\sum_{y=0}^q B_q^y(-1)^yn!\frac{f!}{[f-(q-y)]!}\frac{k!}{(k-y)!}\Big] \, ,
\end{eqnarray}
where $B_j^k=\frac{j!}{k!(j-k)!}$ is the binomial coefficient.  For example, for $\ell=1$, $Per[\Lambda_{\ket{1010},\ket{2000}}]$ is given by
\begin{eqnarray}
Per[\Lambda_{\ket{1010},\ket{2000}}]=(-1)^ka^{3}(g-h) \, .
\end{eqnarray}
When $\ell=2$, $Per_{\Lambda_{\ket{0020},\ket{1001}}}$ can be written as
\begin{eqnarray}
Per[\Lambda_{\ket{0020},\ket{1001}}]=2\cdot(-1)^k a^{4}\left[(f-k)(g-h)\right] \, .
\end{eqnarray}
Then, the probabilities for the configurations $\ket{2000}$  and $\ket{0020}$ are given by
\begin{eqnarray}
P_{\ket{2000}}&&=\frac{1}{4}e^{-2|\alpha|^2}\Big(\frac{|\alpha|^4r_0^2}{2}+|\alpha|^2r_1^2+\frac{r_2^2}{4}+\sqrt{2}|\alpha|^3r_0r_1\cos(\phi_1) \nonumber \\
&&+\frac{\sqrt{2}|\alpha|^2r_0r_2\cos(\phi_2)}{2}+|\alpha|r_1r_2\cos(\phi_1-\phi_2)\Big) \, ,\\
P_{\ket{0020}}&&=\frac{1}{4}e^{-2|\alpha|^2}\Big(\frac{|\alpha|^4r_0^2}{2}+|\alpha|^2r_1^2+\frac{r_2^2}{4}+\sqrt{2}|\alpha|^3r_0r_1\sin(\phi_1) \nonumber \\
&&-\frac{\sqrt{2}|\alpha|^2r_0r_2\cos(\phi_2)}{2}-|\alpha|r_1r_2\sin(\phi_1-\phi_2)\Big) \, .
\end{eqnarray}
For the two-photon configuration ($s=2$), we can code the different phase differences $\phi_j-\phi_k$ and probabilities amplitudes $r_j$ for $j,k\in\{0,1,2\}$.
	
The same process can be considered for each configuration with a fix number of photons $s$, and we can find that the unknown phases and amplitudes will be encoded in the different probabilities. This will produce more complex mathematical expressions, thus being hard to calculate the analytical expression to recover some information from the unknown state~$\ket{\eta_3}$. Nevertheless, it looks a suitable problem for ML methods, which can learn and recognize patterns from data. It is important to mention that when the reference states are quantum superpositions $(\ket{0}+e^{i\theta}\ket{1})/\sqrt{2}$ for $\theta=0$ and $\theta=\pi/2$, the mathematical form of the output probabilities are simple. In this case, analytical expressions for all the amplitudes and phases may be found in a straightforward manner, reaching full tomography with $100\%$ accuracy.

\begin{figure}[t]
\centering
\includegraphics[height=4cm,width=8cm]{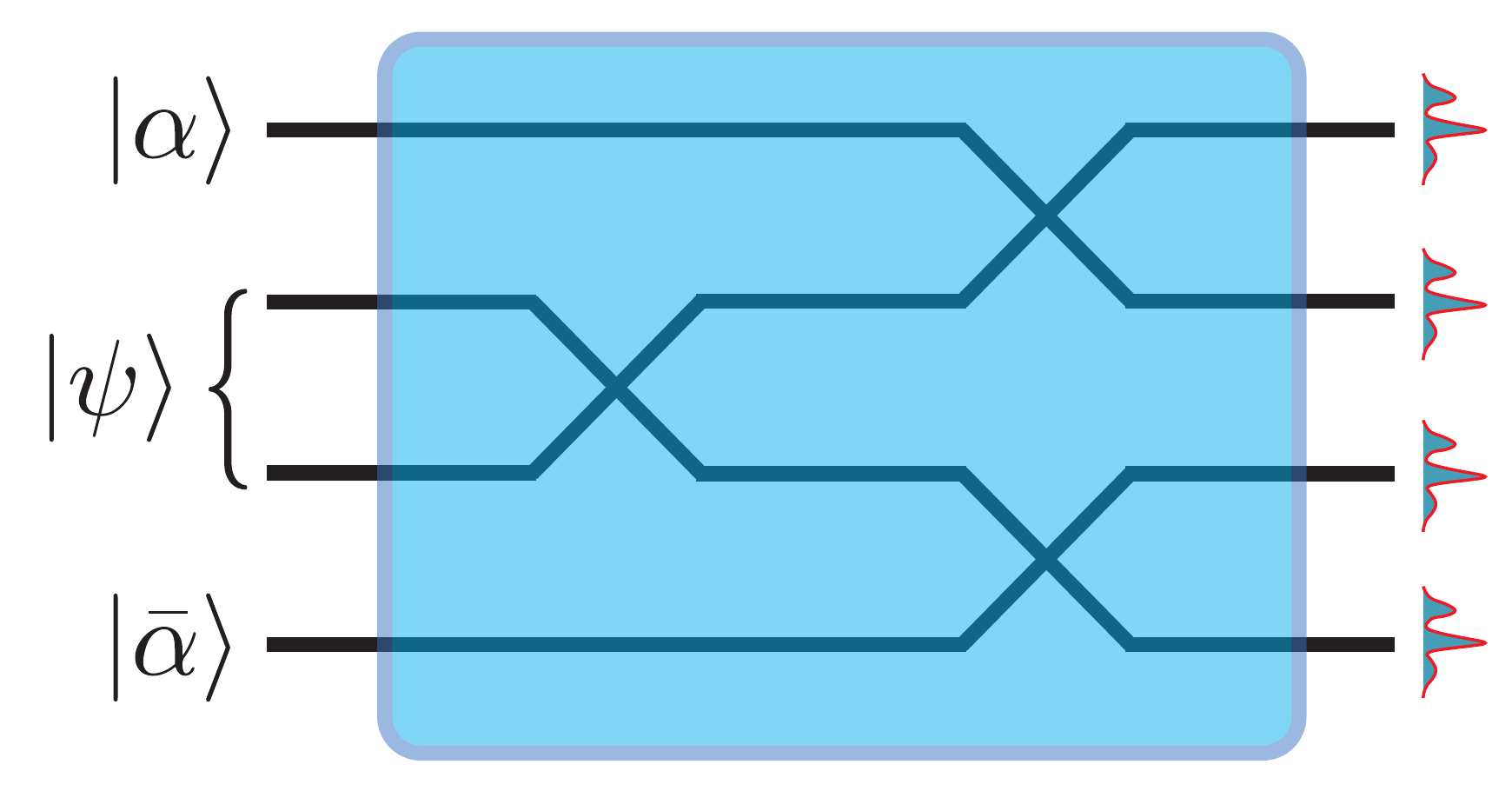}
\caption{Core optical circuit for quantum tomography made up of three beam splitters and the initial input state.}	
\label{entanglement}
\end{figure}

\section{\label{Sec4}Entanglement estimation}

Entanglement is one of the most important resources for quantum information. It describes nonlocal correlations between quantum states, and it has become an important tool for understanding the states of many-body systems. For bipartite systems, the entanglement entropy has become a theoretical measure for categorizing such states. For the entanglement coding, we used the optical circuit shown in Fig.~\ref{entanglement}, where there is an unknown two-mode input state.  We will make use of the von Neumann entropy of a state, which is defined as 
\begin{eqnarray}
S(\rho)=-Tr(\rho\ln\rho),
\label{S}
\end{eqnarray}
where $\rho$ is the density operator of the composite system, while $Tr(\cdot)$ denotes the trace. 

For the calculation of the initial state, we suppose that the first and last modes are coherent states given by
\begin{equation}
|{\alpha_{j}^{(\theta)}}\rangle = \sum_{n=0}^{\infty}\frac{e^{-\frac{1}{2}|\alpha|^2} e^{in\theta}|\alpha|^n}{\sqrt{n!}}|n_j\rangle \, ,
\end{equation}
 with $\theta = 0$ for the mode $1$,  $\theta=\pi/2$ for the mode $4$ as in the previous case,  and a bipartite system for the second and third mode, it means a wave function of the following form
 \begin{equation}
|\psi\rangle = \sum_{j,v=0}^{N}r_{jv}e^{i\phi_{jv}}|j_2\rangle|v_3\rangle \, ,
\end{equation}
where the parameters $r_{jv}$ and $\phi_{jv}$ satisfy $\sum_{j,v=0}^N r_{jv}^2=1$ and $\phi_{jv}\in [0,2\pi]$. Then, the initial state of our optical circuit is
\begin{equation}
\ket{\Psi}=\sum_{m,n=0}^{\infty} \sum_{j,v=0}^{N}i^n\frac{e^{-|\alpha|^2} |\alpha|^{m+n}}{\sqrt{m!n!}}r_{jv} e^{i\phi_{jv}}\ket{m_1}\ket{j_2}\ket{v_3}\ket{n_4} \, .
\label{ini_ent}
\end{equation}

We describe the bipartite system in modes $3$ and $4$ by the density operator $\rho_{AB}=|\psi\rangle \langle\psi|$, and the reduced density matrix of the subsystem $A(B)$ as $\rho_A = Tr_B(\rho_{AB})\, (\rho_B = Tr_A(\rho_{AB}))$. $Tr_{\mathcal{S}}$ means tracing over subsystem $\mathcal{S}$. The entanglement $E(\rho_{AB})$ of a bipartite system $\rho_{AB}$ may be defined as the von Neumann entropy of the reduced density matrix of a subsystem,
\begin{eqnarray}
E(\rho_{AB})=S(\rho_A)=S(\rho_B).
\end{eqnarray}

For simplicity, we rewrite the state given by Eq.~(\ref{ini_ent}) as
\begin{eqnarray}
\ket{\Psi}=\sum_{s=0}^{\infty}|\psi^I_{s}\rangle \, ,
\end{eqnarray}
where $|\psi^I_{s}\rangle$ is a non-normalized state of the superposition of all the four-mode states with $s$ photons, which reads
\begin{eqnarray}
|\psi_{s}^I\rangle = \sum_{j,v=0}^N\sum_{n=0}^{s-j-v}i^n\frac{e^{-|\alpha|^2} |\alpha|^{s-j-v}}{\sqrt{(s-j-v-n)!n!}}r_{jv} e^{i\phi_{jv}} \nonumber \\
\times\ket{(s-j-v-n)_1}\ket{j_2}\ket{v _3}\ket{n_4} \, .
\end{eqnarray}
Then, the output state is a superposition of the different configurations of how photons may have arrived to output modes,
\begin{eqnarray}
|\psi_{s}^O\rangle=\sum_{C_s}\gamma_{C_s}|ghkf\rangle_{C_s},
\end{eqnarray}
where $C_s$ denotes a particular configuration with $s$ photons. The probability of measuring a specific configuration $C_s$ reads
\begin{eqnarray}
P_{C_s} = e^{2|\alpha|^2} \Big|\sum_{j,v=0}^N\sum_{n=0}^{s-j-v} \frac{i^n|\alpha|^{s-j-v}e^{i\phi_{jv}}r_{jv}Per[\Lambda_{\ket{s^*jvn},\ket{ghkf}}]}{(s-j-v-n)!n!\sqrt{g!h!k!f!j!v!}}\Big|^2 , \nonumber \\
\label{prob_ent}
\end{eqnarray}
where $s^*=s-j-v-n$. Now, each set of probabilities is related to a given degree of entanglement. Then, again via the use of a ML protocol based on training a pattern recognition algorithm, we can estimate this essential feature.

\section{\label{Sec5}Machine learning for state characterization}

ML was used to find the relation between the patterns and the initial arbitrary state so that the model, once trained, could infer the state of a new pattern not belonging to the training set. In particular, we developed regression models for the values of amplitude and phase, calculating the associated fidelity as figure-of-merit. Due to the sparsity of the data set, Support Vector Regressors (SVRs) were chosen to carry out the regression of amplitudes and phases as our first approach~\cite{Scholkopf2018MIT, Alvarez_SRep_2017}. SVRs are a generalization of Support Vector Machines, which try to solve classification problems by formulating them as convex optimization problems. In this manner, one has to find the suitable hyperplanes that classify correctly as many training samples as possible. In particular, SVRs work by creating a transformed data space in which the problem is more easily solvable and, ideally the problem is transformed into a linear one. That transformation between spaces is carried out by the so-called kernels. Gaussian, linear, and polynomial kernels have been used in this experimentation. An SVR introduces a region in the hyperspace of the problem called $\epsilon$-tube, within which, all predictions are considered as correct.

The first three tables show the results of the predictions made by the ML models,  evaluating those predictions by means of the fidelity as figure-of-merit. For each data instance, there is  a corresponding fidelity value; the tables show the mean value of the fidelities as well as its standard deviation and the quartiles as measures of dipersal.  A percentile X\% means that in X\% of our data samples, fidelites are below the percentile X\% value.  For example, in Table \ref{tab:fidelity}, for the RBF (Gaussian SVR), 25\% of the samples give fidelities below 0.6397; this of course means that 75\% of the results give fidelity values above 0.6397.
Tables \ref{tab:mae_svr_123}~and~\ref{tab:mae_etr_123} report the entanglement estimation.  The entropy mean value is given for reference purposes to compare its magnitude and its mean absolute error (MAE). We also give the standard deviation of the absolute errors as well as their quartiles and the coefficient of determination ($R^2$-score) for the models.  The coefficient of determination evaluates the quality of the regression, being 1 its best possible value; a constant model that gives the same output value disregarding the input features would have a score of 0.

Starting with the specific results of each experiment,  table~\ref{tab:fidelity} shows the values of fidelity obtained by SVRs implemented with Gaussian (RBF), linear, and polynomial kernels for the case of $N=3$. The maximum fidelities are achieved by the Gaussian kernel, that will be hence selected for ulterior analyses. The polynomial kernel obtains slightly lower fidelities than the Gaussian one, whereas the linear kernel shows the poorest modeling capability among the three. 

\begin{table}[h!] 
\begin{center}
 \begin{tabular}{|c|c|c|c|} 
\hline
 & RBF & Linear & Polynomial \\ [0.5ex] 
\hline
Mean & 0.742530 & 0.655262 & 0.735201 \\ 
Standard Deviation & 0.215096 & 0.237184 & 0.221150 \\
 Minimum & 0.072721 & 0.006989 & 0.063000 \\
25\% & 0.639721 & 0.486128 & 0.620836 \\
50\% & 0.807703 & 0.699080 & 0.802010 \\
75\% & 0.914678 & 0.844586 & 0.926584 \\ 
Maximum & 0.992325 & 0.983482 & 0.991405 \\
\hline
\end{tabular}
\end{center}
\caption{Fidelities achieved by SVRs in the N=3 case with three different kernels: Radial Basis Function (Gaussian), linear, and polynomial. The first column specifies the metrics used to assess the fidelities,namely,  the mean value alongside the corresponding standard deviation as dispersal measure, as well as the minimum,  maximum and quartile values to give information about the distribution of fidelity measures.}
\label{tab:fidelity}
\end{table}

For the sake of comparison with SVRs, we considered a state-of-the-art method, such as Extremely Randomized Trees (ERTs)~\cite{Geurts_ML_2006}, which build multiple regression trees. Each tree takes a random subset of the input features, while nodes are randomly split for the whole data set (in contrast with the well-known Random Forest, there is no bootstrap). To reduce the problem dimensionality, we made use of a principal component analysis (PCA)~\cite{Alpaydin2004MIT}. We set the explained variance ratio at $0.999$, so that we could get a good-enough representation of our data that allows for a reliable fidelity modeling while considerably reducing the computational time.
    
Table \ref{tab:fidelity_svr_123} shows the fidelity values obtained by SVRs with a Gaussian kernel, for non-PCA and PCA versions of the data sets.  The use of PCA seems to have no visible effects on the  fidelity performance. In fact, fidelities tend to be slightly higher when using PCA, thus suggesting its suitability for dimensionality reduction. Table~\ref{tab:fidelity_svr_123} also shows that the most reliable fidelities are achieved for $N=2$, slightly lower values are obtained for $N=1$, and even lower for~$N=3$.

\begin{table}[h!] 
\begin{center}
\begin{tabular}{|c|c|c|c|}
\hline
Without PCA& N=1 & N=2 &  N=3 \\ [0.5ex] 
\hline
Mean &0.823151 & 0.850400 & 0.742530 \\ 
Standard Deviation & 0.190404 &  0.204878 & 0.215096 \\
 Minimum & 0.054745 & 0.105860 & 0.072721 \\
25\% & 0.740702 &  0.787348 &  0.639721 \\
50\% & 0.881910 &  0.950054 & 0.807703 \\
75\% & 0.970362 &   0.983219 & 0.914678 \\ 
Maximum & 0.999827 &  0.998861 & 0.992325 \\
\hline
\end{tabular}
\begin{tabular}{|c|c|c|c|} 
\hline
With PCA & N=1 & N=2 &  N=3 \\ [0.5ex] 
\hline
Mean &0.816223 &   0.867969 &   0.743020 \\
Standard Deviation &0.221873 &   0.200754 &   0.212160 \\
Minimum &0.058625 &   0.127300 &   0.135495 \\
25\% &0.734036 &   0.811426 &   0.636835 \\
50\% &0.908766 &   0.963327 &   0.783768 \\
75\% &0.986301 &   0.991305 &   0.912941 \\
Maximum &0.999982 &   0.998994 &   0.998755 \\
\hline
\end{tabular}
\end{center}
\caption{Fidelities achieved by SVRs with a Gaussian kernel for the three different data sets: N=1, N=2 and N=3.  Results using a preprocessing based on a Principal Component Analysis (PCA) and not using PCA are compared. The first column specifies the metrics used to assess the fidelities,  namely,  the mean value alongside the corresponding standard deviation as dispersal measure, as well as minimum,  maximum and quartile values to give information about the distribution of fidelity measures.}
\label{tab:fidelity_svr_123}
\end{table}

\begin{table}[h!] 
\begin{center}
\begin{tabular}{|c|c|c|c|} 
\hline
Without PCA& N=1 & N=2 &  N=3 \\ [0.5ex] 
\hline
Mean & 0.719850 & 0.900017 &  0.761558 \\ 
Standard Deviation & 0.227072 & 0.191405 &   0.206226\\
Minimum & 0.069204 & 0.138863 & 0.090518\\
25\% & 0.557748 & 0.934672 &  0.671530 \\
50\% & 0.751323 &  0.987036 &   0.813736  \\
75\% & 0.916520 &  0.994959 &  0.928541 \\ 
Maximum & 0.999996 & 0.999894 &  0.988078  \\
\hline
\end{tabular}
\begin{tabular}{|c|c|c|c|} 
\hline
With PCA& N=1 & N=2 &  N=3 \\ [0.5ex] 
\hline
Mean &0.664226 &   0.922877 &   0.787241 \\
Standard Deviation &0.250087 &   0.157877 &   0.189888 \\
Minimum & 0.066888 &   0.050081 &   0.116078 \\
25\% & 0.467351 &   0.947035 &   0.702872 \\
50\% & 0.703973 &   0.981661 &   0.852074 \\
75\% & 0.892137 &   0.992386 &   0.936313 \\
Maximum & 0.999233 &   0.999749 &   0.994355 \\
\hline
\end{tabular}
\end{center}
\caption{Fidelities achieved by ERTs for the three different data sets: N=1, N=2 and N=3.  Results using a preprocessing based on a Principal Component Analysis (PCA) and not using PCA are compared. The first column specifies the metrics used to assess the fidelities,  namely,  the mean value alongside the corresponding standard deviation as dispersal measure, as well as minimum,  maximum and quartile values to give information about the distribution of fidelity measures.}
\label{tab:fidelity_etr_123}
\end{table}
	
Table \ref{tab:fidelity_etr_123} includes the values of fidelity obtained by ERTs for both non-PCA and PCA versions of the data sets.  As in the case of SVRs,  the use of PCA does not affect the  performances in a great extent. In particular, the use of PCA provides slightly better results for $N=2$ and $N=3$, while for $N=1$ the use of the original data without PCA yields higher values of fidelity. The comparison of tables~\ref{tab:fidelity_svr_123} and \ref{tab:fidelity_etr_123} show that the ERTs carry out a better regression than SVRs for $N=2$ and $N=3$, being a SVR approach better for $N=1$. This might likely be due to the fact that $N=1$ corresponds to a more sparse data set, where the SVR is a more adequate and natural solution. Furthermore, that data set may not contain enough variability to exploit ERT capabilities. In summary,  the use of SVRs is suggested for $N=1$ and ERTs for any $N>1$. Furthermore, as PCA does not have a relevant impact on performance, tends to lead to higher fidelities, and allows a reduction of the computational times, its use to reduce the dimensionality of the data sets is encouraged. 
    
Results about entanglement estimation via the corresponding entropy are shown in tables~\ref{tab:mae_svr_123}~and~\ref{tab:mae_etr_123}. The performance of SVRs and ERTs are compared via the mean absolute error (MAE) and $R^2$ scores for different data sets. ERTs outperform SVRs in all cases except for $N=1$, when there is no PCA preprocessing again very likely due to sparsity. This conjecture is reinforced by the fact that after using PCA, hence reducing sparsity, SVRs are not better regressors than ERTs, even for $N=1$.  As in the estimation of fidelities, PCA does not have a relevant impact on performance. In fact, its use tends to lead to slightly better performances (higher R$^2$-scores and lower MAEs) whilst allowing a reduction of the computational times. Therefore, its use to reduce the dimensionality of the data sets is encouraged. We may infer that, for entanglement estimation, ERTs represent a more adequate choice than SVRs. 

\begin{table}[t!] 
\begin{center}
\begin{tabular}{|c|c|c|c|}
\hline
Without PCA& N=1 & N=2 &  N=3 \\ [0.5ex] 
\hline
Entropy Mean Value & 0.297344 & 0.297822 & 0.159738\\
MAE & 0.082243 & 0.142873 & 0.118600 \\ 
Standard Deviation & 0.067844&0.147551&0.121629 \\
Minimum & 0.000213&0.000861&0.000331 \\
25\% & 0.034487&0.065343&0.074025 \\
50\% & 0.072087&0.093323&0.096292 \\
75\% & 0.104777&0.161541&0.115779 \\ 
Maximum & 0.436492&0.934630&1.082693 \\
R$^2$-score & 0.818407 & 0.600526 &  0.507575 \\
\hline
\end{tabular}
\begin{tabular}{|c|c|c|c|} 
\hline
With PCA & N=1 & N=2 &  N=3 \\ [0.5ex] 
\hline
Entropy Mean Value & 0.297344 & 0.297822 & 0.159738 \\
MAE &0.085969&0.132719&0.113547 \\
Standard Deviation &0.071639&0.123191&0.122406 \\
Minimum &0.000968&0.003429&0.002213 \\
25\% &0.038139&0.063135&0.060838 \\
50\% &0.070745&0.095576&0.093087 \\
75\% &0.112728&0.147261&0.113246 \\
Maximum &0.518710&0.654891&1.169980 \\
R$^2$-score & 0.800968 &   0.689208 &   0.522331 \\
\hline
\end{tabular}
\end{center}
\caption{SVRs with Gaussian kernel: entanglement estimation (by entropy) performance for 200 test samples and the three different data sets: N=1, N=2 and N=3.  Results using a preprocessing based on a Principal Component Analysis (PCA) and not using PCA are compared.  The first column specifies the mean values of entropy (as a reference for the committed errors) and the metrics used to evaluate the performance. namely,  Mean Absolute Error (MAE) with its corresponding standard deviation,  as well as the minimum,  quartile and maximum values,  and also R$^2$-scores. }
\label{tab:mae_svr_123}
\end{table}

\section{\label{Sec6}Conclusion}
We have proposed a state characterization via a pattern recognition algorithm in ML for photonics. It is based on the estimation of quantum features by using the output photon number distribution in  a photonic circuit, similar to boson-sampling protocols for continuous variables. We have focused on the estimation of single-mode phases and amplitudes and, also, on the two-mode entanglement estimation. 

SVRs and ERTs have been used to extrapolate the estimation for states not present in the training set, i.e., new probability patterns. The obtained fidelities hint that ML estimations are reliable and can be used to boost the proposed QT protocol.  In particular,  the use of ERTs with previous PCA preprocessing seems to be a suitable approach for setups with~$N>1$. 

The same two ML approaches were employed for entanglement estimation through modeling the von Neumann entropy,  yielding  $R^2$ scores higher than 0.75, thus suggesting the appropriateness of ML for this relevant task.  In this case,  the use of ERTs with a previous PCA preprocessing turns out to be the most adequate choice for all cases.

The proposed photonic circuit with four modes and three beam splitters is experimentally accessible. It means that by using a bigger circuit, with more output probabilities for the number of photons, ML algorithms will likely increase its performance, since the patterns will be more complex and will encode more information.  Nevertheless, the numerical generation of the data will be a hard task due to the complexity of the problem. Also, the generation of experimental data for the training set is a demanding task due to the difficulties to characterize a quantum state.  In any case, it is a useful technique even if we use simple photonic circuits or historical data produced by current photonic experiments for more complex setups, Finally, this work shows that ML techniques can be suitably used for state characterization without the burden of full tomography, paving the way for more sophisticated tools that may help for fast estimation of quantum features.

\begin{table}[h!] 
\begin{center}
\begin{tabular}{|c|c|c|c|}
\hline
Without PCA& N=1 & N=2 &  N=3 \\ [0.5ex] 
\hline
Real Mean & 0.297344 & 0.297822 & 0.159738 \\
MAE &0.081303&0.131039&0.084841 \\ 
Standard Deviation & 0.078961&0.143680&0.108290 \\
Minimum & 0.000018&0.000094&0.000005 \\
25\% & 0.018078&0.019206&0.012170 \\
50\% & 0.054899&0.082159&0.042033 \\
75\% & 0.130161&0.191042&0.117126 \\ 
Maximum & 0.408034&0.704982&0.573232 \\
R$^2$-score & 0.766088 & 0.627655 &  0.664271 \\
\hline
\end{tabular}
\begin{tabular}{|c|c|c|c|} 
\hline
With PCA & N=1 & N=2 &  N=3 \\ [0.5ex] 
\hline
Real Mean & 0.297344 & 0.297822 & 0.159738 \\
MAE &0.073482&0.094449&0.060760 \\
Standard Deviation &0.073704&0.110135&0.103725 \\
Minimum &0.000082&0.000058&0.000004 \\
25\% &0.018621&0.010909&0.007419 \\
50\% &0.047475&0.050592&0.026540 \\
75\% &0.100847&0.138207&0.075164 \\
Maximum &0.386189&0.484120&0.971228 \\
R$^2$-score & 0.822516 &   0.770742 &   0.752039 \\
\hline
\end{tabular}
\end{center}
\caption{ERTs: entanglement estimation (by entropy) performance for 200 test samples and the three different data sets: N=1, N=2 and N=3.  Results using a preprocessing based on a Principal Component Analysis (PCA) and not using PCA are compared.  The first column specifies the mean values of entropy (as a reference for the committed errors) and the metrics used to evaluate the performance. namely,  Mean Absolute Error (MAE) with its corresponding standard deviation,  as well as the minimum,  quartile and maximum values,  and also R$^2$-scores. }
\label{tab:mae_etr_123}
\end{table}

\section{Acknowledgments}
We acknowledge support from QMiCS (820505) and OpenSuperQ (820363) projects from EU Flagship on Quantum Technologies, National Natural Science Foundation of China grant (NSFC) (12075145), Shanghai Government grant STCSM (2019SHZDZX01-ZX04), Spanish Government grant PGC2018-095113-B-I00 (MCIU/AEI/FEDER, UE), Basque Government IT986-16, EU FET Open Quromorphic and EPIQUS projects.

\end{document}